\documentclass[prl,noshowpacs,english,superscriptaddress,twocolumn]{revtex4}
\usepackage{amsmath,amssymb,amsfonts,mathrsfs}
\usepackage{amsthm}
\usepackage{CJK}
\usepackage{bm}
\usepackage{babel}
\usepackage{graphicx}
\allowdisplaybreaks
\usepackage{braket}
\usepackage[breaklinks]{hyperref}
\usepackage{color}
\hypersetup{colorlinks=true, linkcolor=blue, citecolor=blue, filecolor=blue, urlcolor=blue}

\begin{document}

\title{Symmetry-Protected Topological Triangular Weyl Complex}

\author{R. Wang}
\affiliation{Department of Physics $\&$ Institute for Quantum Science and Engineering, Southern University of Science and Technology,
Shenzhen 518055, P. R. China.}
\affiliation{Institute for Structure and Function $\&$
Department of physics $\&$ Center for Quantum Materials and Devices, Chongqing University, Chongqing 400044, P. R. China.}
\author{B. W. Xia}
\affiliation{Department of Physics $\&$ Institute for Quantum Science and Engineering, Southern University of Science and Technology,
Shenzhen 518055, P. R. China.}
\author{Z. J. Chen}
\affiliation{Department of Physics $\&$ Institute for Quantum Science and Engineering, Southern University of Science and Technology,
Shenzhen 518055, P. R. China.}
\affiliation{Department of Physics, South China University of Technology, Guangzhou 510640, P. R. China}
\author{B. B. Zheng}
\affiliation{Department of Physics $\&$ Institute for Quantum Science and Engineering, Southern University of Science and Technology,
Shenzhen 518055, P. R. China.}
\author{Y. J. Zhao}
\affiliation{Department of Physics, South China University of Technology, Guangzhou 510640, P. R. China}
\author{H. Xu}
\email[]{xuh@sustech.edu.cn}
\affiliation{Department of Physics $\&$ Institute for Quantum Science and Engineering, Southern University of Science and Technology,
Shenzhen 518055, P. R. China.}
\affiliation{Guangdong Provincial Key Laboratory of Computational Science and Material Design, Southern University of Science and Technology, Shenzhen 518055, P. R. China.}

\begin{abstract}
Weyl points are often believed to appear in pairs with opposite chirality. In this work, we show by first-principles calculations and symmetry analysis that single Weyl phonons with linear dispersion and double Weyl phonons with quadratic dispersion are simultaneously present between two specific phonon branches in realistic materials with trigonal or hexagonal lattices. These phonon Weyl points are guaranteed to locate at high-symmetry points due to the screw rotational symmetry, forming a unique triangular Weyl complex. In sharp contrast to conventional Weyl systems with surface arcs terminated at the projections of a pair of Weyl points with opposite chirality, the phonon surface arcs of the unconventional triangular Weyl complex connect the projections of one double Weyl point and two single Weyl points. Importantly, the phonon surface arcs originating from the triangular Weyl complex are extremely long and span the entire surface Brillouin-zone. Furthermore, there are only nontrivial phonon surface states across the iso-frequency surface, which facilitates their detection in experiments and further applications. Our work not only offers the promising triangular phonon Weyl complex but also provides guidance for exploring triangular Weyl bosons in both phononic and photonic systems.
\end{abstract}

\pacs{73.20.At, 71.55.Ak, 74.43.-f}

\keywords{ }

\maketitle

Recently, condensed-matter systems with intrinsic topological orders have attracted a lot of attention \cite{Kane-RevModPhys.82.3045,ZSC-RevModPhys.83.1057, RevModPhys.90.015001}. On the one hand, these topological systems provide exotic platforms to study elementary particles and their related phenomena in high-energy physics, since quasiparticle excitations in realistic materials provide analogues of relativistic fermions or bosons in quantum field theory \cite{RevModPhys.90.015001,Wan2011,Weng2015}. On the other hand, topological quasiparticles in crystalline solids arise from nontrivial topology characterized by topological invariants, offering a fascinating avenue to investigate  symmetry-protected topological orders. Furthermore, quasiparticles in crystalline solids are not constrained by the Poincare symmetry but instead of the crystal symmetry. Therefore, beyond conventional Dirac, Weyl, and Majorana particles in the standard model, it is potential to uncover unconventional topological quasiparticles without high-energy physics counterparts in condensed-matter physics \cite{Bradlynaaf5037, ZhuPhysRevX.6.031003, lvNature2017, Winkler_2019}.

Up to now, many conventional and unconventional topological quasiparticles have been proposed. For examples, various nontrivial fermions in topological semimetals \cite{Wan2011,Weng2015, Wang2016prl2, Autes2016, Wang2016prl1, PhysRevB.97.195157} and topological bosons in crystalline solids \cite{PhysRevLett.120.016401, Miao2018, PhysRevB.96.064106, Wangnanoletter, NCR, PhysRevLett.117.068001, PhysRevB.97.054305, PhysRevMaterials.2.114204, PhysRevB.98.220103, PhysRevB.99.174306, PhysRevLett.122.104302} are the subjects of intense studies. Among these nontrivial quasiparticles, Weyl-type excitations are of particular importance. The topology of Weyl point (WP) is characterized by a quantized chiral charge or Chern number $\mathcal{C}$. Due to the twofold-degenerate feature, WPs are present in a system by breaking either the time-reversal ($\mathcal{T}$) or inversion ($\mathcal{I}$) symmetry. Usually, there are equal numbers of WPs with opposite chirality according to the Nielsen-Ninomiya no-go theorem \cite{nogo1, nogo2}, and thus the total topological charge is zero. However, the crystal symmetries of crystalline solids are more complicated, which may possess unconventional Weyl-type quasiparticles. For instance, the 4-fold or 6-fold rotational symmetry can protect quadratic-double or cubic-triple WPs \cite{PhysRevLett.108.266802} and the screw rotational symmetry can protect double WPs \cite{PhysRevLett.120.016401} or WPs with the higher Chern number $\mathcal{C}$ \cite{PhysRevLett.119.206401, PhysRevLett.119.206402}. It is worth noting that the mentioned high-order WPs above always come in pairs with opposite chiral charge \cite{PhysRevLett.108.266802, PhysRevLett.120.016401, PhysRevLett.119.206401, PhysRevLett.119.206402}. If a system simultaneously possesses WPs with different chiral charge, the topological stability (i.e., the conservation of chiral charge) may not require that WPs must appear in pairs \cite{nogo1, nogo2}, i.e., a special Weyl complex with the number of WPs exceeding two can emerge in realistic materials.

\begin{figure}
	\centering
	\includegraphics[scale=0.24]{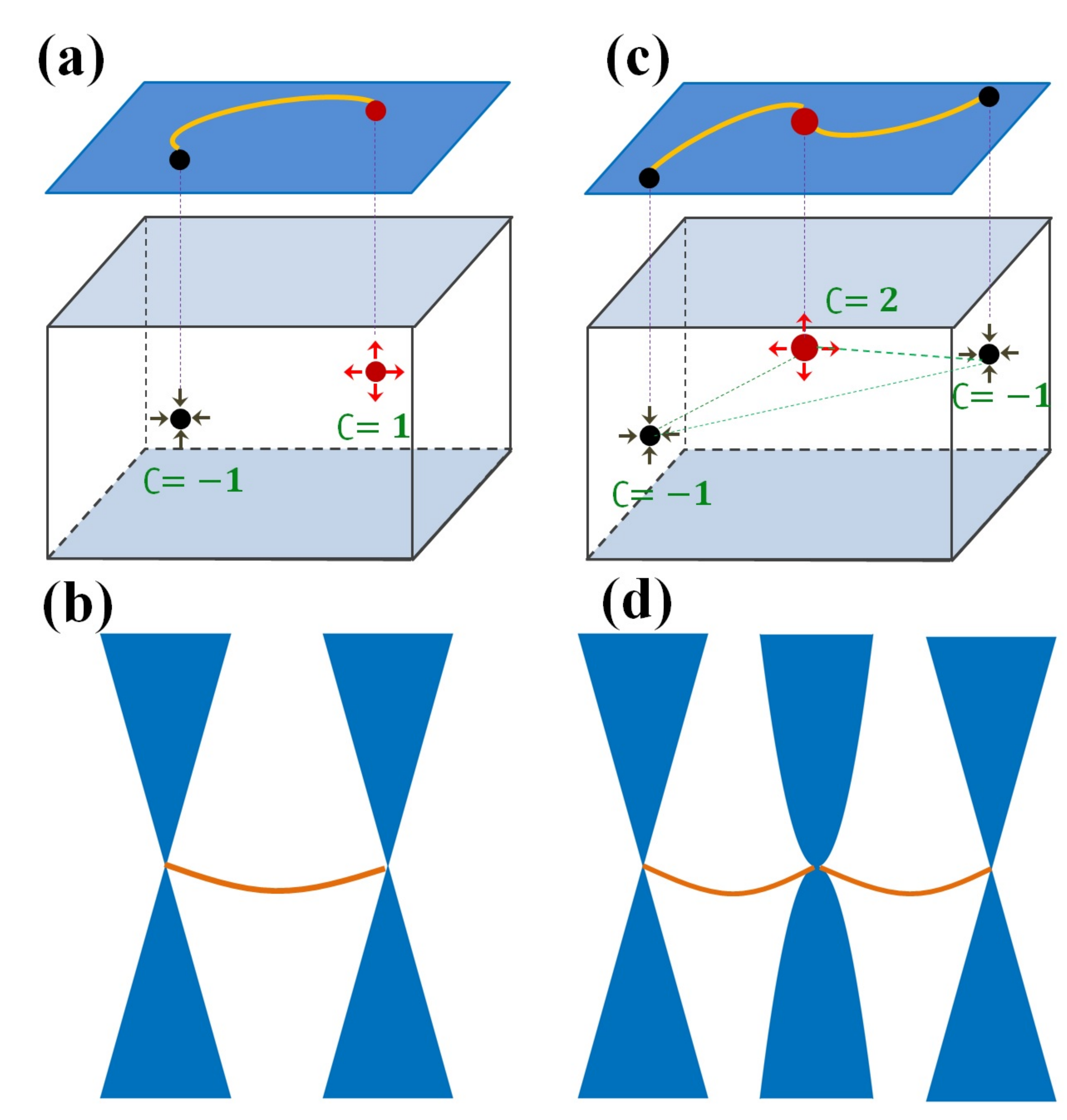}
	\caption{ (a) A Weyl pair with a surface arc and (b) the corresponding surface states. (c) A special Weyl complex (i.e, two single WPs with $\mathcal{C}= -1$ and one double WP with $\mathcal{C}= +2$ form a triangle denoted as green dashed lines) with surface arcs and (d) the corresponding surface states.
\label{figure1}}
\end{figure}

To elaborate the topological features above, we first consider a conventional Weyl system in which WPs appear in pairs with opposite chiral charge as illustrated in Fig. \ref{figure1}(a). Without loss of generality, we assume that the paired WPs are single WPs with Chern number $\mathcal{C}=\pm 1$. This Weyl pair projected on a surface gives a discontinuous surface arc, which connects the projections of these two WPs with opposite chiral charge [see Figs. \ref{figure1}(a) and \ref{figure1}(b)] \cite{Wan2011}. In contrast, if WPs in a system don't appear in pairs with opposite chiral charge, the number of right-handed WPs is unequal to that of left-handed WPs. For instance, two single WPs with Chern number $\mathcal{C}= -1$ and one quadratic-double WP with Chern number $\mathcal{C}= +2$ in a system [see Fig. \ref{figure1}(c)] form a unique triangular Weyl complex. In this case, the topological stability and no-go theorem are also preserved. As a result, this triangular Weyl complex projected on a surface can lead to two surface arcs, which both start at the projection of the double WP and respectively end at the projections of two single WPs [see Figs. \ref{figure1}(c) and \ref{figure1}(d)]. Analogously, if a Weyl system simultaneously possesses WPs with different (e.g., linear, quadratic, cubic) dispersions and Chern numbers, the combination of these WPs can further form various polygonal Weyl complexes. So far, Weyl pairs always appear in a Weyl system \cite{RevModPhys.90.015001}, but Weyl complexes (i.e, triangular or polygonal) with the number of WPs exceeding two and their corresponding surface states have not been reported.

In this work, we identify that unconventional triangular Weyl complex of phonons can exist in crystals crystallizing in trigonal or hexagonal lattices. This corresponds to the fact that the nonsymmorphic screw rotational and $\mathcal{T}$ symmetries are present but the $\mathcal{I}$ symmetry is absent in these space groups. We offer an intuitive perspective of symmetry analysis to understand the symmetry-protected triangular Weyl phonons. For trigonal lattices, we generally consider the nonsymmorphic screw rotational symmetry $\tilde{C}_{3}$ along the $c$-axis, i.e., $\tilde{C}_{3z} = \{C_{3z}|\tau\}$, where $C_{3z}$ is the 3-fold rotational operator and $\tau=(0, 0, \frac{c}{3})$ is a partial translation vector.  In a periodic system, the eigenvalues of $\tilde{C}_{3z}$ can be expressed as $\tilde{E}_{\mu} = E_{\mu} e^{-ik_z c/3}$ (see the Supplemental Material (SM) \cite{SM}), where $E_{\mu}=e^{i2\pi \mu/3}$ ($\mu = 0, 1, 2$) are the rotational eigenvalues of $C_{3z}$.  If two phonon branches are very close in a frequency at $\mathbf{K}_{\mathrm{wp}}$ on a screw axis, we can use a $2\times 2$ effective Hamiltonian to describe them as
\begin{equation}
\mathcal{H}(\mathbf{q})=d(\mathbf{q})\sigma_{+}+d(\mathbf{q})^*\sigma_{-}+f(\mathbf{q})\sigma_z,
\end{equation}
where $\mathcal{H}$ is referenced to the frequency of a WP, $d(\mathbf{q})$ represents a complex function, $f(\mathbf{q})$  represents  a real function, $\mathbf{q} = \mathbf{k} - \mathbf{K}_{\mathrm{wp}}$ denotes the wave vector relative to $\mathbf{K}_{\mathrm{wp}}$,  $\sigma_{\pm}= \sigma_x \pm i\sigma_y$, and $\sigma_{i}$ ($i=x$, $y$, $z$) are the Pauli matrices. $\tilde{C}_{3z}$ constrains the Hamiltonian as
\begin{equation} \label{Hcon}
\tilde{C}_{3z}\mathcal{H}(\mathbf{q})\tilde{C}_{3z}^{-1}=\mathcal{H}(R_{3z}\mathbf{q}),
\end{equation}
where $R_{3z}$ is a $3\times 3$ rotation matrix of $C_{3z}$. On the invariant $k_z$ axis through a high-symmetry point, Eq. (\ref{Hcon}) indicates that all branches on this invariant line correspond to $\tilde{E}_{\mu}$. If two crossing phonon branches are labelled by $\tilde{E}_{\mu_1}$ and $\tilde{E}_{\mu_2}$, the constraint Eq. (\ref{Hcon}) gives
\begin{equation}\label{df}
\begin{split}
e^{-i2\pi(\mu_1 - \mu_2)/3}d(q_{+}, q_{-})&=d(q_{+}e^{i2\pi/3}, q_{-}e^{-i2\pi/3}),\\
f(q_{+}, q_{-})&=f(q_{+}e^{i2\pi/3}, q_{-}e^{-i2\pi/3}),
\end{split}
\end{equation}
where $q_{\pm}=q_x\pm iq_{y}$. The degeneracy of two phonon branches at the invariant plane with $k_z=0, \pm \frac{\pi}{c}$ requires $\mu_1 - \mu_2 = \pm 1$ (see the SM \cite{SM}). At the $K$ (or $H$) point, there is only $\tilde{C}_{3z}$. Then, we expand Eq. (\ref{df}) and remain the lowest orders as
\begin{equation}\label{KH}
d(\mathbf{q})=a_{+}q_+ +a_{-}q_-, \ \ \ f(\mathbf{q})=a_z q_{z},
\end{equation}
which implies a single WP with chiral charge $\pm 1$ at the $K$ (or $H$) point. At the $\Gamma$ (or $A$) point, it is invariant under the $\mathcal{T}$ symmetry. Since the $\mathcal{T}$ symmetry is always conserved in a phonon system, the product operator $\tilde{C}_{3z}\mathcal{T}$ requires (see the SM \cite{SM})
\begin{equation}\label{dft}
\begin{split}
e^{i2\pi(\mu_1 - \mu_2)/3}d(q_{+}, q_{-})&=d(q_{+}e^{-i\pi/3}, q_{-}e^{i\pi/3}),\\
f(q_{+}, q_{-})&=f(q_{+}e^{-i\pi/3}, q_{-}e^{i\pi/3}),
\end{split}
\end{equation}
In this case, the symmetry-allowed expressions as a function of $\mathbf{q}$ to the lowest orders can be written as
\begin{equation}\label{AG}
d(\mathbf{q})=b_{+}q_+^{2}+b_{-}q_-^{2}, \ \ \ f(\mathbf{q})=b_z q_{z},
\end{equation}
which indicates that quasiparticle excitations around the $\Gamma$ (or $A$) point are quadratic in the $k_x$-$k_y$ plane and linear along the $k_z$ axis, forming a double WP with chiral charge $\pm 2$ at the $\Gamma$ (or $A$) point. Besides, the crystals crystallized in a hexagonal lattice possess the $\tilde{C}_{6z}$ symmetry at the $\Gamma$ (or $A$) point and the $\tilde{C}_{3z}$ symmetry at the $K$ (or $H$) point, which can also lead to double and single WPs (see the SM \cite{SM}), respectively.

\begin{figure}
\setlength{\abovecaptionskip}{-0.1cm}
	\centering
	\includegraphics[scale=0.115]{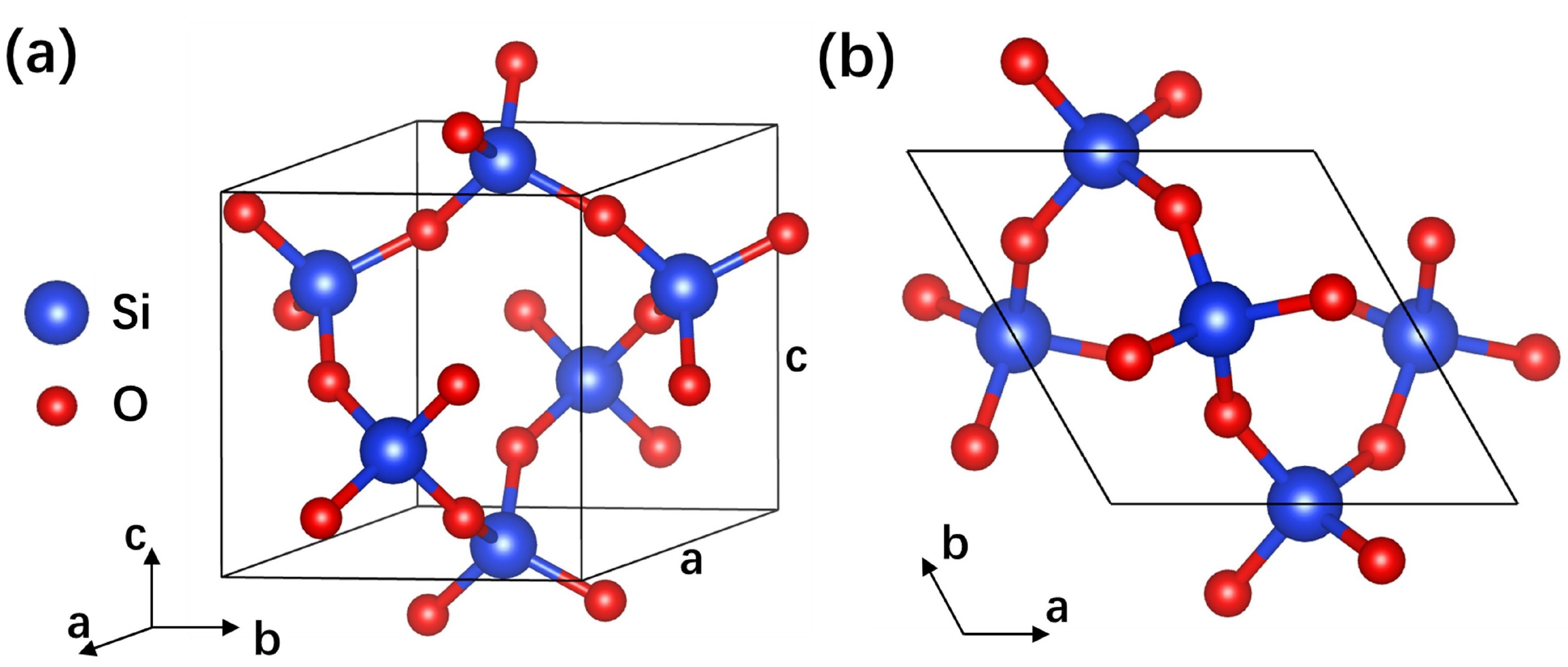}
	\includegraphics[scale=0.45]{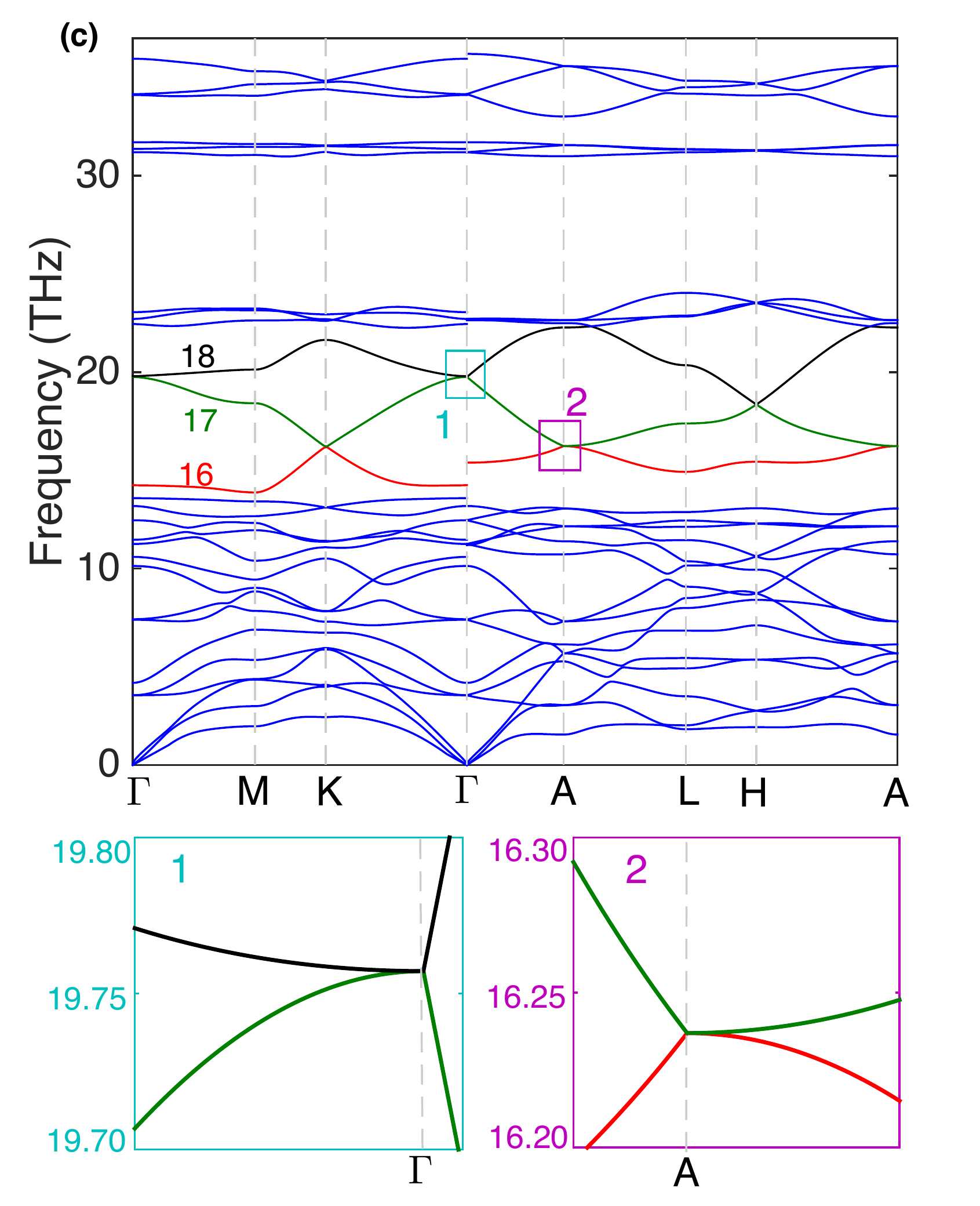}
	\caption{Crystal and phonon spectra of $\alpha$-SiO$_2$. (a) Side and (b) top views of $\alpha$-SiO$_2$. (c) The phonon spectra of $\alpha$-SiO$_2$ along high-symmetry lines. Three nontrivial phonon branches 16, 17, and 18 are highlighted in red, green, and black, respectively. The lower panels show the zoom-in regions 1 and 2 marked by boxes in the upper panel of (c), respectively.
\label{figure2}}
\end{figure}

Based on the above analysis, we have revealed that the combination of $\tilde{C}_{3z}$ and $\mathcal{T}$ symmetries or the $\tilde{C}_{6z}$ symmetry can protect double Weyl phonons at the $\Gamma$ (or $A$) point while the $\tilde{C}_{3z}$ symmetry can protect single Weyl phonons at the $K$ (or $H$) point, resulting in the symmetry-protected triangular Weyl phonons. The possible space groups that host the symmetry-protected triangular Weyl complex of nontrivial phonons are provided in the SM \cite{SM}. In the main text, we focus on $\alpha$-quartz (i.e., $\alpha$-SiO$_2$), a well-known mineral crystallizing in a trigonal lattice. Topological phonon features of YPt$_2$B in a hexagonal lattice are shown in the SM \cite{SM}.

To show the nontrivial phonon topology, we carried out first-principles calculations as implemented in the Vienna $ab$ $initio$ simulation package \cite{Kresse2} (see details in the SM \cite{SM}). The phonon spectra were obtained from a supercell approach, in which interatomic force constants were calculated by finite displacements \cite{Togo2015}. As $\alpha$-SiO$_2$ is polarized, we also considered the nonanalytical term correction to remove imaginary acoustic modes when $\mathbf{q}\rightarrow 0$ \cite{PhysRevB.1.910, PhysRevB.97.094108, PhysRevLett.68.3603}. As shown in Figs. \ref{figure2}(a) and \ref{figure2}(b), $\alpha$-SiO$_2$ crystallizes in a trigonal lattice with nonsymmorphic space group $P3{_2}21$ (No. 154), which lacks the $\mathcal{I}$ symmetry. The hexagonal bulk Brillouin-zone (BZ) and its corresponding (001) and (010) surface BZs are shown Fig. S2 (see the SM \cite{SM}).

\begin{figure}
\setlength{\abovecaptionskip}{-0.4 cm}
	\centering
	\includegraphics[scale=0.33]{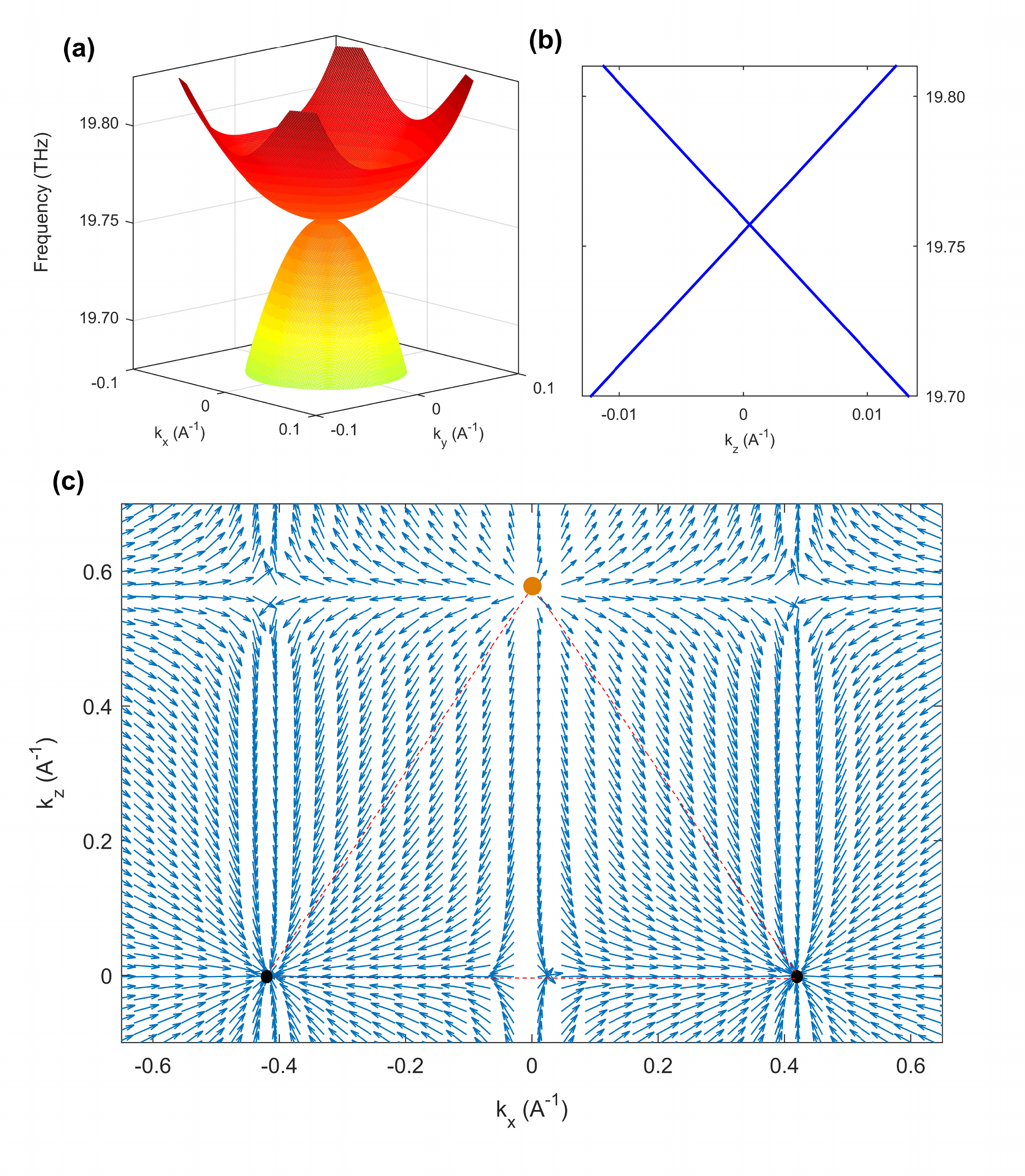}
	\caption{Two phonon branches around the $\Gamma$ point form (a) quadratic dispersion in the $k_x$-$k_y$ plane and (b) linear dispersion along the $k_z$ axis. (c) The distribution of Berry curvature in the $k_x$-$k_z$ plane. The triangular Weyl complex is denoted as the red-dashed triangle.
\label{figure3}}
\end{figure}

Using first-principles calculations, we calculate the phonon spectra and confirm that the triangular Weyl phonons exist in optical phonon branches of $\alpha$-SiO$_2$. The phonon dispersion curves along the high-symmetry directions are shown in Fig. \ref{figure2}(c), which match well with previous theoretical and experimental results \cite{PhysRevB.97.224306, PhysRevB.50.13035, Swainson, PhysRevB.73.094304, Dorner_1980}. The phonon spectra show that there are visible double-degenerate nodal points at high-symmetry points, which are contributed from three phonon branches 16, 17, and 18 [highlighted in Fig. \ref{figure2}(c)].  The branches 16 and 17 cross at the $K$ and $A$ points. Instead, the branches 17 and 18 cross at the $H$  and $\Gamma$ points.  The crossing frequencies at the $K$, $A$,  $H$, and $\Gamma$ points are $\omega_K = 16.13$ THz, $\omega_A = 16.23$ THz, $\omega_H = 18.35$ THz, and $\omega_{\Gamma} = 19.76$ THz, respectively. Around the $K$ (or $H$) point, it is clear to see that phonon dispersions are linear. This isotropic Dirac cone indicates the presence of a single WP with chiral charge $\pm 1$. In contrast, phonon dispersions around the $\Gamma$ (or $A$) point are dramatically different; that is, quadratically dispersing along $\Gamma$-$K$ (or $A$-$H$) and linearly dispersing along $\Gamma$-$A$.  We plot phonon dispersions around the $\Gamma$ point in the $k_x$-$k_y$ plane with $k_z = 0$ and along the $k_z$ axis in Figs. \ref{figure3}(a) and \ref{figure3}(b), respectively. The two figures show that the phonon dispersions around the $\Gamma$  point are quadratic in the $k_x$-$k_y$ plane but linear along the $k_z$ axis. We also check the phonon dispersions around the $A$ point, which are the same with those around $\Gamma$.  These results indicate that the nodal point at $\Gamma$ (or $A$) form a double phonon WP with chiral charge $-2$ (or  $+2$).

We further employ the Wilson-loop method \cite{WU2017,Yu2011} to determine the chiral charge of the above Weyl phonons.  The results show that the WPs at the $K$ and $A$ points between the branches 16 and 17 are indeed the single WP with $\mathcal{C} = -1$ and double WP with $\mathcal{C} = +2$, respectively. There are six $K$ points in the first BZ and each $K$ point is shared by three neighbor Wigner-Seitz cells in momentum space, and thus the total chiral charge is zero.  As shown in Fig. \ref{figure3}(c), we plot the corresponding distribution of Berry curvature in the $k_x$-$k_z$ plane. As expected, the double phonon WP with $\mathcal{C}= +2$ at the $A$ point acts as the ``source" point, whereas the single phonon WPs with $\mathcal{C}=-1$ at the $K$ point can be viewed as the ``sink" points. This unique distribution of phonon WPs in momentum space forms the triangular Weyl complex. Similarly, the phonon WPs between the branches 17 and 18 possess the opposite chiral charge (i.e., six single WPs with $\mathcal{C} = +1$ at the $H$ point and one double WP with $\mathcal{C} = -2$ at the $\Gamma$ point), which also indicates the other triangular Weyl phonons but exhibits the opposite distribution of Berry curvature.

\begin{figure}
	\centering
	\includegraphics[scale=0.4]{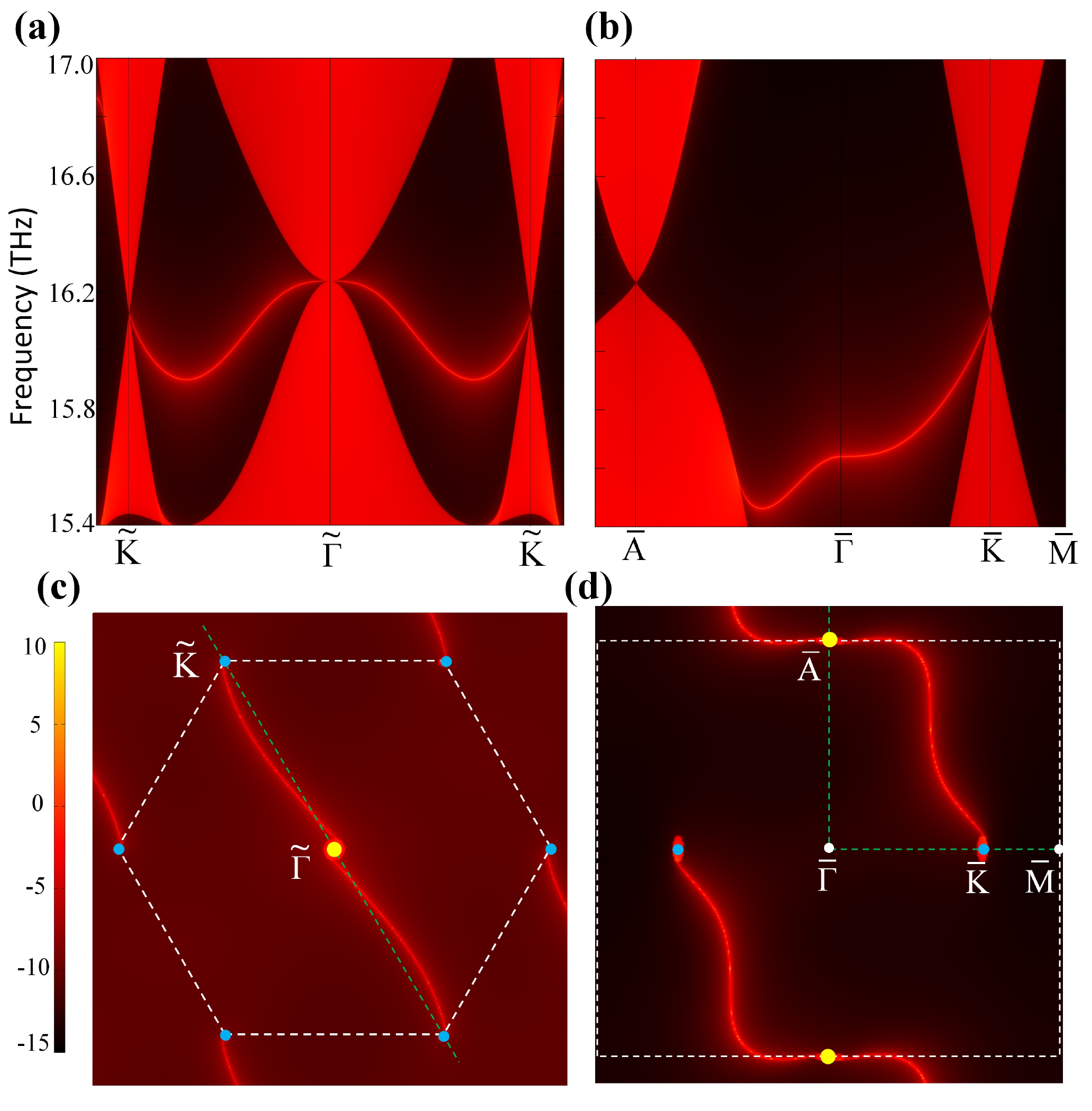}
	\caption{The phonon surface states of $\alpha$-SiO$_{2}$. The phonon LDOS projected on the (a) (001) surface and (b) (010) surface. The iso-frequency surfaces at $\omega = 16.22$ THz projected on the (c) (001) and (d) (010) surfaces. In (c) and (d), the first BZ of (001) and (010) are marked by white-dashed lines, and high-symmetry lines for LDOS in (a) and (b) are marked in green-dashed lines. The yellow and blue dots indicate the projections of WPs with $\mathcal{C}= +2$ and $\mathcal{C}=-1$, respectively.
\label{figure4}}
\end{figure}

The exotic triangular Weyl complex of phonons corresponds to unique nontrivial surface states. To illustrate this, we construct a phonon tight-binding Hamiltonian in the Wannier representation from second-order interatomic force constants. In this representation, the iterative Green's function method \cite{WU2017, Sancho1984} is employed to calculate phonon surface states.  The phonon local density of states (LDOS) projected on a semi-infinite (001) surface of $\alpha$-SiO$_{2}$ is shown in Fig. \ref{figure4}(a).  As expected, there are two visible phonon surface states, which both start at the projection of the double WP at $\tilde{\Gamma}$ and respectively end at the projections of two single WPs at $\tilde{K}$. Unlike the surface arcs terminated at the projections of a pair of WPs with opposite chiral charge in conventional Weyl systems, the iso-frequency surface of (001) shows that there are two phonon surface arcs connecting the projections of the double WP and single WPs in the first surface BZ of (001) [see Fig. \ref{figure4}(c)]. Due to the conservation of topological chiral charge, the other phonon surface arcs terminated at $\tilde{K}$ are shared by the neighboring Wigner-Seitz cells. We also plot the phonon LDOS and iso-frequency surface projected on the semi-infinite (010) surface of $\alpha$-SiO$_{2}$ in Figs. \ref{figure4}(b) and \ref{figure4}(d), respectively. The projections of bulk states on the (010) surface confirm that the dispersion near the double WP along the $k_z$ axis is linear. Two phonon surface arcs in the first BZ of the (010) surface are clearly visible. Since the symmetry-protected WPs at the $A$ ($\Gamma$) and $K$ ($H$) points dominate the iso-frequency surface of 3D hexagonal BZ, we can see that two phonon surface arcs cross each half of the first surface BZ. Therefore, the phonon surface arcs are guaranteed to be very long and span the entire first surface BZ. Furthermore, it is worth mentioning that there are only the nontrivial surface arcs across the iso-frequency surface. The absence of trivial bulk states on the (001) and (010) surfaces of $\alpha$-SiO$_{2}$ greatly facilitates the experimental detection and further applications.

In summary, we show that there are unequal numbers of phonon WPs with opposite chirality in trigonal or hexagonal lattices with breaking the inversion symmetry. The distribution of Weyl phonons constructs an unconventional triangular Weyl complex, and their nontrivial phonon surface states uniquely connect the projections of phonon WPs with different chiral charge. The nonsymmorphic screw symmetry protects the phonon WPs located at high-symmetry points, which guarantees that the phonon surface arcs span the entire first BZ of surface.  The longest phonon arcs can provide entire modes of topological phonon surface states in a robust nontrivial one-way phonon propagation channel. A very interesting point is that trivial bulk states are absent in iso-frequency surfaces, which greatly facilitates their detection in experiments. Therefore, our findings provide ideal candidates for realizing triangular Weyl complexes of phonons and their nontrivial surface states. Furthermore, our results can also be applied to fermionic systems.

This work is supported by the National Natural Science Foundation of China (NSFC, Grant Nos.11674148, 11974062, 11974160, and 11947406), the Guangdong Natural Science Funds for Distinguished Young Scholars (No. 2017B030306008), the Fundamental Research Funds for the Central Universities of China (No. 2019CDXYWL0029), the Chongqing Natural Science Funds (No. cstc2019jcyj-msxmX0563), the fund of the Guangdong Provincial Key Laboratory of Computational Science and Material Design (No.2019B030301001), and the Center for Computational Science and Engineering of Southern University of Science and Technology.


\begin{thebibliography}{43}
\expandafter\ifx\csname natexlab\endcsname\relax\def\natexlab#1{#1}\fi
\expandafter\ifx\csname bibnamefont\endcsname\relax
  \def\bibnamefont#1{#1}\fi
\expandafter\ifx\csname bibfnamefont\endcsname\relax
  \def\bibfnamefont#1{#1}\fi
\expandafter\ifx\csname citenamefont\endcsname\relax
  \def\citenamefont#1{#1}\fi
\expandafter\ifx\csname url\endcsname\relax
  \def\url#1{\texttt{#1}}\fi
\expandafter\ifx\csname urlprefix\endcsname\relax\def\urlprefix{URL }\fi
\providecommand{\bibinfo}[2]{#2}
\providecommand{\eprint}[2][]{\url{#2}}

\bibitem[{\citenamefont{Hasan and Kane}(2010)}]{Kane-RevModPhys.82.3045}
\bibinfo{author}{\bibfnamefont{M.~Z.} \bibnamefont{Hasan}} \bibnamefont{and}
  \bibinfo{author}{\bibfnamefont{C.~L.} \bibnamefont{Kane}},
  \bibinfo{journal}{Rev. Mod. Phys.} \textbf{\bibinfo{volume}{82}},
  \bibinfo{pages}{3045} (\bibinfo{year}{2010}).

\bibitem[{\citenamefont{Qi and Zhang}(2011)}]{ZSC-RevModPhys.83.1057}
\bibinfo{author}{\bibfnamefont{X.-L.} \bibnamefont{Qi}} \bibnamefont{and}
  \bibinfo{author}{\bibfnamefont{S.-C.} \bibnamefont{Zhang}},
  \bibinfo{journal}{Rev. Mod. Phys.} \textbf{\bibinfo{volume}{83}},
  \bibinfo{pages}{1057} (\bibinfo{year}{2011}).

\bibitem[{\citenamefont{Armitage et~al.}(2018)\citenamefont{Armitage, Mele, and
  Vishwanath}}]{RevModPhys.90.015001}
\bibinfo{author}{\bibfnamefont{N.~P.} \bibnamefont{Armitage}},
  \bibinfo{author}{\bibfnamefont{E.~J.} \bibnamefont{Mele}}, \bibnamefont{and}
  \bibinfo{author}{\bibfnamefont{A.}~\bibnamefont{Vishwanath}},
  \bibinfo{journal}{Rev. Mod. Phys.} \textbf{\bibinfo{volume}{90}},
  \bibinfo{pages}{015001} (\bibinfo{year}{2018}).

\bibitem[{\citenamefont{Wan et~al.}(2011)\citenamefont{Wan, Turner, Vishwanath,
  and Savrasov}}]{Wan2011}
\bibinfo{author}{\bibfnamefont{X.}~\bibnamefont{Wan}},
  \bibinfo{author}{\bibfnamefont{A.~M.} \bibnamefont{Turner}},
  \bibinfo{author}{\bibfnamefont{A.}~\bibnamefont{Vishwanath}},
  \bibnamefont{and} \bibinfo{author}{\bibfnamefont{S.~Y.}
  \bibnamefont{Savrasov}}, \bibinfo{journal}{Phys. Rev. B}
  \textbf{\bibinfo{volume}{83}}, \bibinfo{pages}{205101}
  (\bibinfo{year}{2011}).

\bibitem[{\citenamefont{Weng et~al.}(2015)\citenamefont{Weng, Fang, Fang,
  Bernevig, and Dai}}]{Weng2015}
\bibinfo{author}{\bibfnamefont{H.}~\bibnamefont{Weng}},
  \bibinfo{author}{\bibfnamefont{C.}~\bibnamefont{Fang}},
  \bibinfo{author}{\bibfnamefont{Z.}~\bibnamefont{Fang}},
  \bibinfo{author}{\bibfnamefont{B.~A.} \bibnamefont{Bernevig}},
  \bibnamefont{and} \bibinfo{author}{\bibfnamefont{X.}~\bibnamefont{Dai}},
  \bibinfo{journal}{Phys. Rev. X} \textbf{\bibinfo{volume}{5}},
  \bibinfo{pages}{011029} (\bibinfo{year}{2015}).

\bibitem[{\citenamefont{Bradlyn et~al.}(2016)\citenamefont{Bradlyn, Cano, Wang,
  Vergniory, Felser, Cava, and Bernevig}}]{Bradlynaaf5037}
\bibinfo{author}{\bibfnamefont{B.}~\bibnamefont{Bradlyn}},
  \bibinfo{author}{\bibfnamefont{J.}~\bibnamefont{Cano}},
  \bibinfo{author}{\bibfnamefont{Z.}~\bibnamefont{Wang}},
  \bibinfo{author}{\bibfnamefont{M.~G.} \bibnamefont{Vergniory}},
  \bibinfo{author}{\bibfnamefont{C.}~\bibnamefont{Felser}},
  \bibinfo{author}{\bibfnamefont{R.~J.} \bibnamefont{Cava}}, \bibnamefont{and}
  \bibinfo{author}{\bibfnamefont{B.~A.} \bibnamefont{Bernevig}},
  \bibinfo{journal}{Science} \textbf{\bibinfo{volume}{353}},
  \bibinfo{pages}{6299} (\bibinfo{year}{2016}).

\bibitem[{\citenamefont{Zhu et~al.}(2016)\citenamefont{Zhu, Winkler, Wu, Li,
  and Soluyanov}}]{ZhuPhysRevX.6.031003}
\bibinfo{author}{\bibfnamefont{Z.}~\bibnamefont{Zhu}},
  \bibinfo{author}{\bibfnamefont{G.~W.} \bibnamefont{Winkler}},
  \bibinfo{author}{\bibfnamefont{Q.}~\bibnamefont{Wu}},
  \bibinfo{author}{\bibfnamefont{J.}~\bibnamefont{Li}}, \bibnamefont{and}
  \bibinfo{author}{\bibfnamefont{A.~A.} \bibnamefont{Soluyanov}},
  \bibinfo{journal}{Phys. Rev. X} \textbf{\bibinfo{volume}{6}},
  \bibinfo{pages}{031003} (\bibinfo{year}{2016}).

\bibitem[{\citenamefont{Lv et~al.}(2017)\citenamefont{Lv, Feng, Xu, Gao, Ma,
  Kong, Richard, Huang, Strocov, Fang et~al.}}]{lvNature2017}
\bibinfo{author}{\bibfnamefont{B.~Q.} \bibnamefont{Lv}},
  \bibinfo{author}{\bibfnamefont{Z.-L.} \bibnamefont{Feng}},
  \bibinfo{author}{\bibfnamefont{Q.-N.} \bibnamefont{Xu}},
  \bibinfo{author}{\bibfnamefont{X.}~\bibnamefont{Gao}},
  \bibinfo{author}{\bibfnamefont{J.-Z.} \bibnamefont{Ma}},
  \bibinfo{author}{\bibfnamefont{L.-Y.} \bibnamefont{Kong}},
  \bibinfo{author}{\bibfnamefont{P.}~\bibnamefont{Richard}},
  \bibinfo{author}{\bibfnamefont{Y.-B.} \bibnamefont{Huang}},
  \bibinfo{author}{\bibfnamefont{V.~N.} \bibnamefont{Strocov}},
  \bibinfo{author}{\bibfnamefont{C.}~\bibnamefont{Fang}}, \bibnamefont{et~al.},
  \bibinfo{journal}{Nature} \textbf{\bibinfo{volume}{546}},
  \bibinfo{pages}{627} (\bibinfo{year}{2017}).

\bibitem[{\citenamefont{Winkler et~al.}(2019)\citenamefont{Winkler, Singh, and
  Soluyanov}}]{Winkler_2019}
\bibinfo{author}{\bibfnamefont{G.~W.} \bibnamefont{Winkler}},
  \bibinfo{author}{\bibfnamefont{S.}~\bibnamefont{Singh}}, \bibnamefont{and}
  \bibinfo{author}{\bibfnamefont{A.~A.} \bibnamefont{Soluyanov}},
  \bibinfo{journal}{Chin. Phys. B} \textbf{\bibinfo{volume}{28}},
  \bibinfo{pages}{077303} (\bibinfo{year}{2019}).

\bibitem[{\citenamefont{Wang et~al.}(2016{\natexlab{a}})\citenamefont{Wang,
  Gresch, Soluyanov, Xie, Kushwaha, Dai, Troyer, Cava, and
  Bernevig}}]{Wang2016prl2}
\bibinfo{author}{\bibfnamefont{Z.}~\bibnamefont{Wang}},
  \bibinfo{author}{\bibfnamefont{D.}~\bibnamefont{Gresch}},
  \bibinfo{author}{\bibfnamefont{A.~A.} \bibnamefont{Soluyanov}},
  \bibinfo{author}{\bibfnamefont{W.}~\bibnamefont{Xie}},
  \bibinfo{author}{\bibfnamefont{S.}~\bibnamefont{Kushwaha}},
  \bibinfo{author}{\bibfnamefont{X.}~\bibnamefont{Dai}},
  \bibinfo{author}{\bibfnamefont{M.}~\bibnamefont{Troyer}},
  \bibinfo{author}{\bibfnamefont{R.~J.} \bibnamefont{Cava}}, \bibnamefont{and}
  \bibinfo{author}{\bibfnamefont{B.~A.} \bibnamefont{Bernevig}},
  \bibinfo{journal}{Phys. Rev. Lett.} \textbf{\bibinfo{volume}{117}},
  \bibinfo{pages}{056805} (\bibinfo{year}{2016}{\natexlab{a}}).

\bibitem[{\citenamefont{Aut\`es et~al.}(2016)\citenamefont{Aut\`es, Gresch,
  Troyer, Soluyanov, and Yazyev}}]{Autes2016}
\bibinfo{author}{\bibfnamefont{G.}~\bibnamefont{Aut\`es}},
  \bibinfo{author}{\bibfnamefont{D.}~\bibnamefont{Gresch}},
  \bibinfo{author}{\bibfnamefont{M.}~\bibnamefont{Troyer}},
  \bibinfo{author}{\bibfnamefont{A.~A.} \bibnamefont{Soluyanov}},
  \bibnamefont{and} \bibinfo{author}{\bibfnamefont{O.~V.}
  \bibnamefont{Yazyev}}, \bibinfo{journal}{Phys. Rev. Lett.}
  \textbf{\bibinfo{volume}{117}}, \bibinfo{pages}{066402}
  (\bibinfo{year}{2016}).

\bibitem[{\citenamefont{Wang et~al.}(2016{\natexlab{b}})\citenamefont{Wang,
  Vergniory, Kushwaha, Hirschberger, Chulkov, Ernst, Ong, Cava, and
  Bernevig}}]{Wang2016prl1}
\bibinfo{author}{\bibfnamefont{Z.}~\bibnamefont{Wang}},
  \bibinfo{author}{\bibfnamefont{M.~G.} \bibnamefont{Vergniory}},
  \bibinfo{author}{\bibfnamefont{S.}~\bibnamefont{Kushwaha}},
  \bibinfo{author}{\bibfnamefont{M.}~\bibnamefont{Hirschberger}},
  \bibinfo{author}{\bibfnamefont{E.~V.} \bibnamefont{Chulkov}},
  \bibinfo{author}{\bibfnamefont{A.}~\bibnamefont{Ernst}},
  \bibinfo{author}{\bibfnamefont{N.~P.} \bibnamefont{Ong}},
  \bibinfo{author}{\bibfnamefont{R.~J.} \bibnamefont{Cava}}, \bibnamefont{and}
  \bibinfo{author}{\bibfnamefont{B.~A.} \bibnamefont{Bernevig}},
  \bibinfo{journal}{Phys. Rev. Lett.} \textbf{\bibinfo{volume}{117}},
  \bibinfo{pages}{236401} (\bibinfo{year}{2016}{\natexlab{b}}).

\bibitem[{\citenamefont{Wang et~al.}(2018)\citenamefont{Wang, Jin, Zhao, Chen,
  Zhao, and Xu}}]{PhysRevB.97.195157}
\bibinfo{author}{\bibfnamefont{R.}~\bibnamefont{Wang}},
  \bibinfo{author}{\bibfnamefont{Y.~J.} \bibnamefont{Jin}},
  \bibinfo{author}{\bibfnamefont{J.~Z.} \bibnamefont{Zhao}},
  \bibinfo{author}{\bibfnamefont{Z.~J.} \bibnamefont{Chen}},
  \bibinfo{author}{\bibfnamefont{Y.~J.} \bibnamefont{Zhao}}, \bibnamefont{and}
  \bibinfo{author}{\bibfnamefont{H.}~\bibnamefont{Xu}}, \bibinfo{journal}{Phys.
  Rev. B} \textbf{\bibinfo{volume}{97}}, \bibinfo{pages}{195157}
  (\bibinfo{year}{2018}).

\bibitem[{\citenamefont{Zhang et~al.}(2018)\citenamefont{Zhang, Song,
  Alexandradinata, Weng, Fang, Lu, and Fang}}]{PhysRevLett.120.016401}
\bibinfo{author}{\bibfnamefont{T.}~\bibnamefont{Zhang}},
  \bibinfo{author}{\bibfnamefont{Z.}~\bibnamefont{Song}},
  \bibinfo{author}{\bibfnamefont{A.}~\bibnamefont{Alexandradinata}},
  \bibinfo{author}{\bibfnamefont{H.}~\bibnamefont{Weng}},
  \bibinfo{author}{\bibfnamefont{C.}~\bibnamefont{Fang}},
  \bibinfo{author}{\bibfnamefont{L.}~\bibnamefont{Lu}}, \bibnamefont{and}
  \bibinfo{author}{\bibfnamefont{Z.}~\bibnamefont{Fang}},
  \bibinfo{journal}{Phys. Rev. Lett.} \textbf{\bibinfo{volume}{120}},
  \bibinfo{pages}{016401} (\bibinfo{year}{2018}).

\bibitem[{\citenamefont{Miao et~al.}(2018)\citenamefont{Miao, Zhang, Wang,
  Meyers, Said, Wang, Shi, Weng, Fang, and Dean}}]{Miao2018}
\bibinfo{author}{\bibfnamefont{H.}~\bibnamefont{Miao}},
  \bibinfo{author}{\bibfnamefont{T.~T.} \bibnamefont{Zhang}},
  \bibinfo{author}{\bibfnamefont{L.}~\bibnamefont{Wang}},
  \bibinfo{author}{\bibfnamefont{D.}~\bibnamefont{Meyers}},
  \bibinfo{author}{\bibfnamefont{A.~H.} \bibnamefont{Said}},
  \bibinfo{author}{\bibfnamefont{Y.~L.} \bibnamefont{Wang}},
  \bibinfo{author}{\bibfnamefont{Y.~G.} \bibnamefont{Shi}},
  \bibinfo{author}{\bibfnamefont{H.~M.} \bibnamefont{Weng}},
  \bibinfo{author}{\bibfnamefont{Z.}~\bibnamefont{Fang}}, \bibnamefont{and}
  \bibinfo{author}{\bibfnamefont{M.~P.~M.} \bibnamefont{Dean}},
  \bibinfo{journal}{Phys. Rev. Lett.} \textbf{\bibinfo{volume}{121}},
  \bibinfo{pages}{035302} (\bibinfo{year}{2018}).

\bibitem[{\citenamefont{Liu et~al.}(2017)\citenamefont{Liu, Xu, Zhang, and
  Duan}}]{PhysRevB.96.064106}
\bibinfo{author}{\bibfnamefont{Y.}~\bibnamefont{Liu}},
  \bibinfo{author}{\bibfnamefont{Y.}~\bibnamefont{Xu}},
  \bibinfo{author}{\bibfnamefont{S.-C.} \bibnamefont{Zhang}}, \bibnamefont{and}
  \bibinfo{author}{\bibfnamefont{W.}~\bibnamefont{Duan}},
  \bibinfo{journal}{Phys. Rev. B} \textbf{\bibinfo{volume}{96}},
  \bibinfo{pages}{064106} (\bibinfo{year}{2017}).

\bibitem[{\citenamefont{Jin et~al.}(2018{\natexlab{a}})\citenamefont{Jin, Wang,
  and Xu}}]{Wangnanoletter}
\bibinfo{author}{\bibfnamefont{Y.}~\bibnamefont{Jin}},
  \bibinfo{author}{\bibfnamefont{R.}~\bibnamefont{Wang}}, \bibnamefont{and}
  \bibinfo{author}{\bibfnamefont{H.}~\bibnamefont{Xu}}, \bibinfo{journal}{Nano
  Lett.} \textbf{\bibinfo{volume}{18}}, \bibinfo{pages}{7755}
  (\bibinfo{year}{2018}{\natexlab{a}}).

\bibitem[{\citenamefont{Liu et~al.}(2018)\citenamefont{Liu, Xu, and
  Duan}}]{NCR}
\bibinfo{author}{\bibfnamefont{Y.}~\bibnamefont{Liu}},
  \bibinfo{author}{\bibfnamefont{Y.}~\bibnamefont{Xu}}, \bibnamefont{and}
  \bibinfo{author}{\bibfnamefont{W.}~\bibnamefont{Duan}},
  \bibinfo{journal}{Nat. Sci. Rev.} \textbf{\bibinfo{volume}{5}},
  \bibinfo{pages}{314} (\bibinfo{year}{2018}).

\bibitem[{\citenamefont{Stenull et~al.}(2016)\citenamefont{Stenull, Kane, and
  Lubensky}}]{PhysRevLett.117.068001}
\bibinfo{author}{\bibfnamefont{O.}~\bibnamefont{Stenull}},
  \bibinfo{author}{\bibfnamefont{C.~L.} \bibnamefont{Kane}}, \bibnamefont{and}
  \bibinfo{author}{\bibfnamefont{T.~C.} \bibnamefont{Lubensky}},
  \bibinfo{journal}{Phys. Rev. Lett.} \textbf{\bibinfo{volume}{117}},
  \bibinfo{pages}{068001} (\bibinfo{year}{2016}).

\bibitem[{\citenamefont{Li et~al.}(2018)\citenamefont{Li, Xie, Ullah, Li, Ma,
  Li, Li, and Chen}}]{PhysRevB.97.054305}
\bibinfo{author}{\bibfnamefont{J.}~\bibnamefont{Li}},
  \bibinfo{author}{\bibfnamefont{Q.}~\bibnamefont{Xie}},
  \bibinfo{author}{\bibfnamefont{S.}~\bibnamefont{Ullah}},
  \bibinfo{author}{\bibfnamefont{R.}~\bibnamefont{Li}},
  \bibinfo{author}{\bibfnamefont{H.}~\bibnamefont{Ma}},
  \bibinfo{author}{\bibfnamefont{D.}~\bibnamefont{Li}},
  \bibinfo{author}{\bibfnamefont{Y.}~\bibnamefont{Li}}, \bibnamefont{and}
  \bibinfo{author}{\bibfnamefont{X.-Q.} \bibnamefont{Chen}},
  \bibinfo{journal}{Phys. Rev. B} \textbf{\bibinfo{volume}{97}},
  \bibinfo{pages}{054305} (\bibinfo{year}{2018}).

\bibitem[{\citenamefont{Singh et~al.}(2018)\citenamefont{Singh, Wu, Yue,
  Romero, and Soluyanov}}]{PhysRevMaterials.2.114204}
\bibinfo{author}{\bibfnamefont{S.}~\bibnamefont{Singh}},
  \bibinfo{author}{\bibfnamefont{Q.}~\bibnamefont{Wu}},
  \bibinfo{author}{\bibfnamefont{C.}~\bibnamefont{Yue}},
  \bibinfo{author}{\bibfnamefont{A.~H.} \bibnamefont{Romero}},
  \bibnamefont{and} \bibinfo{author}{\bibfnamefont{A.~A.}
  \bibnamefont{Soluyanov}}, \bibinfo{journal}{Phys. Rev. Materials}
  \textbf{\bibinfo{volume}{2}}, \bibinfo{pages}{114204} (\bibinfo{year}{2018}).

\bibitem[{\citenamefont{Jin et~al.}(2018{\natexlab{b}})\citenamefont{Jin, Chen,
  Xia, Zhao, Wang, and Xu}}]{PhysRevB.98.220103}
\bibinfo{author}{\bibfnamefont{Y.~J.} \bibnamefont{Jin}},
  \bibinfo{author}{\bibfnamefont{Z.~J.} \bibnamefont{Chen}},
  \bibinfo{author}{\bibfnamefont{B.~W.} \bibnamefont{Xia}},
  \bibinfo{author}{\bibfnamefont{Y.~J.} \bibnamefont{Zhao}},
  \bibinfo{author}{\bibfnamefont{R.}~\bibnamefont{Wang}}, \bibnamefont{and}
  \bibinfo{author}{\bibfnamefont{H.}~\bibnamefont{Xu}}, \bibinfo{journal}{Phys.
  Rev. B} \textbf{\bibinfo{volume}{98}}, \bibinfo{pages}{220103}
  (\bibinfo{year}{2018}{\natexlab{b}}).

\bibitem[{\citenamefont{Xie et~al.}(2019{\natexlab{a}})\citenamefont{Xie, Li,
  Ullah, Li, Wang, Li, Li, Yunoki, and Chen}}]{PhysRevB.99.174306}
\bibinfo{author}{\bibfnamefont{Q.}~\bibnamefont{Xie}},
  \bibinfo{author}{\bibfnamefont{J.}~\bibnamefont{Li}},
  \bibinfo{author}{\bibfnamefont{S.}~\bibnamefont{Ullah}},
  \bibinfo{author}{\bibfnamefont{R.}~\bibnamefont{Li}},
  \bibinfo{author}{\bibfnamefont{L.}~\bibnamefont{Wang}},
  \bibinfo{author}{\bibfnamefont{D.}~\bibnamefont{Li}},
  \bibinfo{author}{\bibfnamefont{Y.}~\bibnamefont{Li}},
  \bibinfo{author}{\bibfnamefont{S.}~\bibnamefont{Yunoki}}, \bibnamefont{and}
  \bibinfo{author}{\bibfnamefont{X.-Q.} \bibnamefont{Chen}},
  \bibinfo{journal}{Phys. Rev. B} \textbf{\bibinfo{volume}{99}},
  \bibinfo{pages}{174306} (\bibinfo{year}{2019}{\natexlab{a}}).

\bibitem[{\citenamefont{Xie et~al.}(2019{\natexlab{b}})\citenamefont{Xie, Liu,
  Cheng, Liu, Chen, and Tian}}]{PhysRevLett.122.104302}
\bibinfo{author}{\bibfnamefont{B.}~\bibnamefont{Xie}},
  \bibinfo{author}{\bibfnamefont{H.}~\bibnamefont{Liu}},
  \bibinfo{author}{\bibfnamefont{H.}~\bibnamefont{Cheng}},
  \bibinfo{author}{\bibfnamefont{Z.}~\bibnamefont{Liu}},
  \bibinfo{author}{\bibfnamefont{S.}~\bibnamefont{Chen}}, \bibnamefont{and}
  \bibinfo{author}{\bibfnamefont{J.}~\bibnamefont{Tian}},
  \bibinfo{journal}{Phys. Rev. Lett.} \textbf{\bibinfo{volume}{122}},
  \bibinfo{pages}{104302} (\bibinfo{year}{2019}{\natexlab{b}}).

\bibitem[{\citenamefont{Nielsen and Ninomiya}(1981{\natexlab{a}})}]{nogo1}
\bibinfo{author}{\bibfnamefont{H.~B.} \bibnamefont{Nielsen}} \bibnamefont{and}
  \bibinfo{author}{\bibfnamefont{M.}~\bibnamefont{Ninomiya}},
  \bibinfo{journal}{Nucl. Phys. B} \textbf{\bibinfo{volume}{185}},
  \bibinfo{pages}{20} (\bibinfo{year}{1981}{\natexlab{a}}).

\bibitem[{\citenamefont{Nielsen and Ninomiya}(1981{\natexlab{b}})}]{nogo2}
\bibinfo{author}{\bibfnamefont{H.~B.} \bibnamefont{Nielsen}} \bibnamefont{and}
  \bibinfo{author}{\bibfnamefont{M.}~\bibnamefont{Ninomiya}},
  \bibinfo{journal}{Nucl. Phys. B} \textbf{\bibinfo{volume}{193}},
  \bibinfo{pages}{173} (\bibinfo{year}{1981}{\natexlab{b}}).

\bibitem[{\citenamefont{Fang et~al.}(2012)\citenamefont{Fang, Gilbert, Dai, and
  Bernevig}}]{PhysRevLett.108.266802}
\bibinfo{author}{\bibfnamefont{C.}~\bibnamefont{Fang}},
  \bibinfo{author}{\bibfnamefont{M.~J.} \bibnamefont{Gilbert}},
  \bibinfo{author}{\bibfnamefont{X.}~\bibnamefont{Dai}}, \bibnamefont{and}
  \bibinfo{author}{\bibfnamefont{B.~A.} \bibnamefont{Bernevig}},
  \bibinfo{journal}{Phys. Rev. Lett.} \textbf{\bibinfo{volume}{108}},
  \bibinfo{pages}{266802} (\bibinfo{year}{2012}).

\bibitem[{\citenamefont{Chang et~al.}(2017)\citenamefont{Chang, Xu, Wieder,
  Sanchez, Huang, Belopolski, Chang, Zhang, Bansil, Lin
  et~al.}}]{PhysRevLett.119.206401}
\bibinfo{author}{\bibfnamefont{G.}~\bibnamefont{Chang}},
  \bibinfo{author}{\bibfnamefont{S.-Y.} \bibnamefont{Xu}},
  \bibinfo{author}{\bibfnamefont{B.~J.} \bibnamefont{Wieder}},
  \bibinfo{author}{\bibfnamefont{D.~S.} \bibnamefont{Sanchez}},
  \bibinfo{author}{\bibfnamefont{S.-M.} \bibnamefont{Huang}},
  \bibinfo{author}{\bibfnamefont{I.}~\bibnamefont{Belopolski}},
  \bibinfo{author}{\bibfnamefont{T.-R.} \bibnamefont{Chang}},
  \bibinfo{author}{\bibfnamefont{S.}~\bibnamefont{Zhang}},
  \bibinfo{author}{\bibfnamefont{A.}~\bibnamefont{Bansil}},
  \bibinfo{author}{\bibfnamefont{H.}~\bibnamefont{Lin}}, \bibnamefont{et~al.},
  \bibinfo{journal}{Phys. Rev. Lett.} \textbf{\bibinfo{volume}{119}},
  \bibinfo{pages}{206401} (\bibinfo{year}{2017}).

\bibitem[{\citenamefont{Tang et~al.}(2017)\citenamefont{Tang, Zhou, and
  Zhang}}]{PhysRevLett.119.206402}
\bibinfo{author}{\bibfnamefont{P.}~\bibnamefont{Tang}},
  \bibinfo{author}{\bibfnamefont{Q.}~\bibnamefont{Zhou}}, \bibnamefont{and}
  \bibinfo{author}{\bibfnamefont{S.-C.} \bibnamefont{Zhang}},
  \bibinfo{journal}{Phys. Rev. Lett.} \textbf{\bibinfo{volume}{119}},
  \bibinfo{pages}{206402} (\bibinfo{year}{2017}).

\bibitem[{SM()}]{SM}
\bibinfo{note}{See Supplemental Material at [url] for the computional method, the detailed symmetry and effective model analysis, the summary of space groups with the triangular Weyl complex, and the results of YPt$_2$B, which inlcudes Refs. \cite{Kohn,Perdew1,Perdew2,Kresse4,Ceperley1980,Monkhorst,DIAS200762}}.


\bibitem[{\citenamefont{Kohn and Sham}(1965)}]{Kohn}
\bibinfo{author}{\bibfnamefont{W.}~\bibnamefont{Kohn}} \bibnamefont{and}
  \bibinfo{author}{\bibfnamefont{L.~J.} \bibnamefont{Sham}},
  \bibinfo{journal}{Phys. Rev.} \textbf{\bibinfo{volume}{140}},
  \bibinfo{pages}{A1133} (\bibinfo{year}{1965}).

\bibitem[{\citenamefont{Perdew et~al.}(1996)\citenamefont{Perdew, Burke, and
  Ernzerhof}}]{Perdew1}
\bibinfo{author}{\bibfnamefont{J.~P.} \bibnamefont{Perdew}},
  \bibinfo{author}{\bibfnamefont{K.}~\bibnamefont{Burke}}, \bibnamefont{and}
  \bibinfo{author}{\bibfnamefont{M.}~\bibnamefont{Ernzerhof}},
  \bibinfo{journal}{Phys. Rev. Lett.} \textbf{\bibinfo{volume}{77}},
  \bibinfo{pages}{3865} (\bibinfo{year}{1996}).

\bibitem[{\citenamefont{Perdew et~al.}(1997)\citenamefont{Perdew, Burke, and
  Ernzerhof}}]{Perdew2}
\bibinfo{author}{\bibfnamefont{J.~P.} \bibnamefont{Perdew}},
  \bibinfo{author}{\bibfnamefont{K.}~\bibnamefont{Burke}}, \bibnamefont{and}
  \bibinfo{author}{\bibfnamefont{M.}~\bibnamefont{Ernzerhof}},
  \bibinfo{journal}{Phys. Rev. Lett.} \textbf{\bibinfo{volume}{78}},
  \bibinfo{pages}{1396} (\bibinfo{year}{1997}).

\bibitem[{\citenamefont{Kresse and Joubert}(1999)}]{Kresse4}
\bibinfo{author}{\bibfnamefont{G.}~\bibnamefont{Kresse}} \bibnamefont{and}
  \bibinfo{author}{\bibfnamefont{D.}~\bibnamefont{Joubert}},
  \bibinfo{journal}{Phys. Rev. B} \textbf{\bibinfo{volume}{59}},
  \bibinfo{pages}{1758} (\bibinfo{year}{1999}).

\bibitem[{\citenamefont{Ceperley and Alder}(1980)}]{Ceperley1980}
\bibinfo{author}{\bibfnamefont{D.~M.} \bibnamefont{Ceperley}} \bibnamefont{and}
  \bibinfo{author}{\bibfnamefont{B.~J.} \bibnamefont{Alder}},
  \bibinfo{journal}{Phys. Rev. Lett.} \textbf{\bibinfo{volume}{45}},
  \bibinfo{pages}{566} (\bibinfo{year}{1980}).

\bibitem[{\citenamefont{Monkhorst and Pack}(1976)}]{Monkhorst}
\bibinfo{author}{\bibfnamefont{H.~J.} \bibnamefont{Monkhorst}}
  \bibnamefont{and} \bibinfo{author}{\bibfnamefont{J.~D.} \bibnamefont{Pack}},
  \bibinfo{journal}{Phys. Rev. B} \textbf{\bibinfo{volume}{13}},
  \bibinfo{pages}{5188} (\bibinfo{year}{1976}).




\bibitem[{\citenamefont{Dias et~al.}(2007)\citenamefont{Dias, Sologub, Pereira,
  and Gon?alves}}]{DIAS200762}
\bibinfo{author}{\bibfnamefont{M.}~\bibnamefont{Dias}},
  \bibinfo{author}{\bibfnamefont{O.}~\bibnamefont{Sologub}},
  \bibinfo{author}{\bibfnamefont{L.}~\bibnamefont{Pereira}}, \bibnamefont{and}
  \bibinfo{author}{\bibfnamefont{A.}~\bibnamefont{Gon\c{c}alves}},
  \bibinfo{journal}{J. Alloy. Comp.} \textbf{\bibinfo{volume}{438}},
  \bibinfo{pages}{62 } (\bibinfo{year}{2007}).


\bibitem[{\citenamefont{Kresse and Furthm{\"{u}}ller}(1996)}]{Kresse2}
\bibinfo{author}{\bibfnamefont{G.}~\bibnamefont{Kresse}} \bibnamefont{and}
  \bibinfo{author}{\bibfnamefont{J.}~\bibnamefont{Furthm{\"{u}}ller}},
  \bibinfo{journal}{Phys. Rev. B} \textbf{\bibinfo{volume}{54}},
  \bibinfo{pages}{11169} (\bibinfo{year}{1996}).

\bibitem[{\citenamefont{Togo and Tanaka}(2015)}]{Togo2015}
\bibinfo{author}{\bibfnamefont{A.}~\bibnamefont{Togo}} \bibnamefont{and}
  \bibinfo{author}{\bibfnamefont{I.}~\bibnamefont{Tanaka}},
  \bibinfo{journal}{Scr. Mater.} \textbf{\bibinfo{volume}{108}},
  \bibinfo{pages}{1} (\bibinfo{year}{2015}).

\bibitem[{\citenamefont{Pick et~al.}(1970)\citenamefont{Pick, Cohen, and
  Martin}}]{PhysRevB.1.910}
\bibinfo{author}{\bibfnamefont{R.~M.} \bibnamefont{Pick}},
  \bibinfo{author}{\bibfnamefont{M.~H.} \bibnamefont{Cohen}}, \bibnamefont{and}
  \bibinfo{author}{\bibfnamefont{R.~M.} \bibnamefont{Martin}},
  \bibinfo{journal}{Phys. Rev. B} \textbf{\bibinfo{volume}{1}},
  \bibinfo{pages}{910} (\bibinfo{year}{1970}).

\bibitem[{\citenamefont{Winta et~al.}(2018)\citenamefont{Winta, Gewinner,
  Sch\"ollkopf, Wolf, and Paarmann}}]{PhysRevB.97.094108}
\bibinfo{author}{\bibfnamefont{C.~J.} \bibnamefont{Winta}},
  \bibinfo{author}{\bibfnamefont{S.}~\bibnamefont{Gewinner}},
  \bibinfo{author}{\bibfnamefont{W.}~\bibnamefont{Sch\"ollkopf}},
  \bibinfo{author}{\bibfnamefont{M.}~\bibnamefont{Wolf}}, \bibnamefont{and}
  \bibinfo{author}{\bibfnamefont{A.}~\bibnamefont{Paarmann}},
  \bibinfo{journal}{Phys. Rev. B} \textbf{\bibinfo{volume}{97}},
  \bibinfo{pages}{094108} (\bibinfo{year}{2018}).

\bibitem[{\citenamefont{Gonze et~al.}(1992)\citenamefont{Gonze, Allan, and
  Teter}}]{PhysRevLett.68.3603}
\bibinfo{author}{\bibfnamefont{X.}~\bibnamefont{Gonze}},
  \bibinfo{author}{\bibfnamefont{D.~C.} \bibnamefont{Allan}}, \bibnamefont{and}
  \bibinfo{author}{\bibfnamefont{M.~P.} \bibnamefont{Teter}},
  \bibinfo{journal}{Phys. Rev. Lett.} \textbf{\bibinfo{volume}{68}},
  \bibinfo{pages}{3603} (\bibinfo{year}{1992}).

\bibitem[{\citenamefont{Mizokami et~al.}(2018)\citenamefont{Mizokami, Togo, and
  Tanaka}}]{PhysRevB.97.224306}
\bibinfo{author}{\bibfnamefont{K.}~\bibnamefont{Mizokami}},
  \bibinfo{author}{\bibfnamefont{A.}~\bibnamefont{Togo}}, \bibnamefont{and}
  \bibinfo{author}{\bibfnamefont{I.}~\bibnamefont{Tanaka}},
  \bibinfo{journal}{Phys. Rev. B} \textbf{\bibinfo{volume}{97}},
  \bibinfo{pages}{224306} (\bibinfo{year}{2018}).

\bibitem[{\citenamefont{Gonze et~al.}(1994)\citenamefont{Gonze, Charlier,
  Allan, and Teter}}]{PhysRevB.50.13035}
\bibinfo{author}{\bibfnamefont{X.}~\bibnamefont{Gonze}},
  \bibinfo{author}{\bibfnamefont{J.-C.} \bibnamefont{Charlier}},
  \bibinfo{author}{\bibfnamefont{D.}~\bibnamefont{Allan}}, \bibnamefont{and}
  \bibinfo{author}{\bibfnamefont{M.}~\bibnamefont{Teter}},
  \bibinfo{journal}{Phys. Rev. B} \textbf{\bibinfo{volume}{50}},
  \bibinfo{pages}{13035} (\bibinfo{year}{1994}).

\bibitem[{\citenamefont{Swainson and Dove}(1995)}]{Swainson}
\bibinfo{author}{\bibfnamefont{I.~P.} \bibnamefont{Swainson}} \bibnamefont{and}
  \bibinfo{author}{\bibfnamefont{M.~T.} \bibnamefont{Dove}},
  \bibinfo{journal}{J. Phys.: Condensed Matter} \textbf{\bibinfo{volume}{7}},
  \bibinfo{pages}{1771} (\bibinfo{year}{1995}).

\bibitem[{\citenamefont{Choudhury and Chaplot}(2006)}]{PhysRevB.73.094304}
\bibinfo{author}{\bibfnamefont{N.}~\bibnamefont{Choudhury}} \bibnamefont{and}
  \bibinfo{author}{\bibfnamefont{S.~L.} \bibnamefont{Chaplot}},
  \bibinfo{journal}{Phys. Rev. B} \textbf{\bibinfo{volume}{73}},
  \bibinfo{pages}{094304} (\bibinfo{year}{2006}).

\bibitem[{\citenamefont{Dorner et~al.}(1980)\citenamefont{Dorner, Grimm, and
  Rzany}}]{Dorner_1980}
\bibinfo{author}{\bibfnamefont{B.}~\bibnamefont{Dorner}},
  \bibinfo{author}{\bibfnamefont{H.}~\bibnamefont{Grimm}}, \bibnamefont{and}
  \bibinfo{author}{\bibfnamefont{H.}~\bibnamefont{Rzany}}, \bibinfo{journal}{J.
  Phys. C: Solid State Phys.} \textbf{\bibinfo{volume}{13}},
  \bibinfo{pages}{6607} (\bibinfo{year}{1980}).

\bibitem[{\citenamefont{Wu et~al.}(2018)\citenamefont{Wu, Zhang, Song, Troyer,
  and Soluyanov}}]{WU2017}
\bibinfo{author}{\bibfnamefont{Q.}~\bibnamefont{Wu}},
  \bibinfo{author}{\bibfnamefont{S.}~\bibnamefont{Zhang}},
  \bibinfo{author}{\bibfnamefont{H.-F.} \bibnamefont{Song}},
  \bibinfo{author}{\bibfnamefont{M.}~\bibnamefont{Troyer}}, \bibnamefont{and}
  \bibinfo{author}{\bibfnamefont{A.~A.} \bibnamefont{Soluyanov}},
  \bibinfo{journal}{Comput. Phys. Commun.} \textbf{\bibinfo{volume}{224}},
  \bibinfo{pages}{405 } (\bibinfo{year}{2018}).

\bibitem[{\citenamefont{Yu et~al.}(2011)\citenamefont{Yu, Qi, Bernevig, Fang,
  and Dai}}]{Yu2011}
\bibinfo{author}{\bibfnamefont{R.}~\bibnamefont{Yu}},
  \bibinfo{author}{\bibfnamefont{X.~L.} \bibnamefont{Qi}},
  \bibinfo{author}{\bibfnamefont{A.}~\bibnamefont{Bernevig}},
  \bibinfo{author}{\bibfnamefont{Z.}~\bibnamefont{Fang}}, \bibnamefont{and}
  \bibinfo{author}{\bibfnamefont{X.}~\bibnamefont{Dai}},
  \bibinfo{journal}{Phys. Rev. B} \textbf{\bibinfo{volume}{84}},
  \bibinfo{pages}{075119} (\bibinfo{year}{2011}).

\bibitem[{\citenamefont{Sancho et~al.}(1984)\citenamefont{Sancho, Sancho, and
  Rubio}}]{Sancho1984}
\bibinfo{author}{\bibfnamefont{M.~P.~L.} \bibnamefont{Sancho}},
  \bibinfo{author}{\bibfnamefont{J.~M.~L.} \bibnamefont{Sancho}},
  \bibnamefont{and} \bibinfo{author}{\bibfnamefont{J.}~\bibnamefont{Rubio}},
  \bibinfo{journal}{J. Phys. F: Metal Phys.} \textbf{\bibinfo{volume}{14}},
  \bibinfo{pages}{1205} (\bibinfo{year}{1984}).

\end{thebibliography}

\end{document}